\documentclass[preprint]{aastex631}

\usepackage{upgreek}
\usepackage{amsmath,amsthm,amssymb}
\usepackage{mathdots} % para el comando \iddots
\usepackage{mathrsfs} % para formato de letra
\usepackage{subfigure}
\begin{document}

\title{Spatio temporal analysis of waves in compressively driven magnetohydrodynamics turbulence}

\author{M. Brodiano}
\affiliation{Instituto de F\'isica de Buenos Aires, CONICET-UBA, Ciudad Universitaria, 1428, Buenos Aires, Argentina}

\author[0000-0002-1272-2778]{N. Andr\'es}
\affiliation{Instituto de Astronomía y Física del Espacio, CONICET-UBA, Ciudad Universitaria, 1428, Buenos Aires,Argentina}
\affiliation{Departamento de F\'isica, Facultad de Ciencias Exactas y Naturales, Universidad de Buenos Aires, Ciudad Universitaria, 1428, Buenos Aires, Argentina.}

\author{P. Dmitruk}
\affiliation{Instituto de F\'isica de Buenos Aires, CONICET-UBA, Ciudad Universitaria, 1428 Buenos Aires, Argentina}
\affiliation{Departamento de F\'isica, Facultad de Ciencias Exactas y Naturales, Universidad de Buenos Aires, Ciudad Universitaria, 1428, Buenos Aires, Argentina.}

\begin{abstract}
Using direct numerical simulations (DNSs), the interaction between linear waves and turbulence under the compressible magnetohydrodynamic (CMHD) approach was studied. A set of DNSs in three dimensions for a spatial resolution of $128^3$ and $256^3$ were performed. A parametric study was carried out varying the sonic Mach number, the mean magnetic field and the compressibility amplitude of the forcing. Spatio-temporal spectra of the magnetic energy were built and analyzed, allowing for direct identification of all wave modes in a CMHD turbulent system and quantification of the amount of energy in each mode as a function of the wave number. Thus, linear waves were detected, that is Alfvén waves and fast and slow magnetosonic waves. Furthermore, different responses of the plasma were found according to whether the Mach number or the mean magnetic field was varied. On the other hand, making use of spatio-temporal spectra and two different integration methods, we accurately quantified the amount of energy present in each of the normal modes. Finally, although the presence of linear waves was observed, in all the cases studied the system was mainly dominated by the non-linear dynamics of the plasma.

\end{abstract}

\section{Introduction} 
\label{sec:intro}

Plasma turbulence is ubiquitous in space and astrophysical flows. For example, the solar wind emitting from the sun into interplanetary space, one of the most studied natural plasma, is in a turbulent state \citep{M2011, bruno2013, Matthaeus2021}. Plasma turbulence is a multiscale phenomenon, involving structures across a wide range of spatial and temporal scales. In the fluid description of plasmas, say magnetohydrodynamics (MHD) theory, only large scales are resolved. For instance, incompressible MHD (IMHD) has a wide range of applications, including those of relevance for planetary science, astrophysics, and nuclear fusion science \citep{priest2012,pouquet1993,Biskamp}. However, this model is inadequate in those media that show significant density fluctuations, such as the ionized interstellar medium, some regions of the solar wind and the earth's magnetosheath \citep{S2020, bruno2013, Armstrong, bavassano1982}. Recent \textit{in situ} observations have shown that compressibility plays a significant role in the turbulent dynamics of the fast and slow solar wind, in particular, by supplying the energy dissipation needed to account for the local heating and particle acceleration of the solar wind \citep{G2016,B2016c,H2017a,H2017b,A2019,A2021}. In particular, recently it has been evidenced in the Earth’s magnetosheath that density fluctuations reinforce the anisotropy of the energy cascade rate with respect to the local magnetic field \citet{H2017b} and increase the cascade rate as it enters into the sub-ion scales \citep{A2019}. Therefore, a study of a compressible MHD (CMHD) turbulence is essential for a deep understanding of the turbulent dynamics of the solar wind at scales larger than the ion inertial length. 

In the presence of a uniform mean magnetic field $\textbf{B}_0$, both the IMHD and CMHD models support the existence of linear waves \citep{Dmi2009}. The IMHD model has Alfvén waves as exact non-linear solutions. These transverse and incompressible waves propagate along the $\textbf{B}_0$ direction. For the CMHD approach, we have to add two propagating compressible wave modes that are not present in the IMHD model, namely fast and slow magnetosonic waves.  When a turbulent regime develops in the presence of waves and eddies, two different regimes can be identified depending on the strength of the non-linear coupling, the so-called weak and strong turbulent regimes. In the first case, this regime can be analytically described by a classical technique based on the perturbation theory to obtain a prediction for the scaling of
the energy spectrum \citep{z1965,GA2000,GA2003,nazarenko2011}. In the second case, waves and structures or vortices coexist with strong coupling, and phenomenological models are often used to study the non-linear dynamics of turbulent plasmas \citep{IR1964,KR1965,Hig1984,Goldreich1995}. Note however that, even in this case, some exact laws, e.g., the so-called $4/5$ law of homogeneous turbulence and its generalizations, can be derived for different fluid approximations of magnetized plasmas \citep{karman1938,Ch1951,P1998a,Go2008,Ga2011,B2013,A2017b}. 

The existence in IMHD of multiple times scales gives rise to multiple phenomenological models of IMHD turbulence \citep{Zhou2004}. In the Iroshnikov-Kraichnan (IK) phenomenology \citep{IR1964,KR1965}, the interaction between waves and eddies results in a quenching of the energy transfer towards small scales, which are assumed to be isotropic. This results in a modification of the Kolmogorov energy spectrum from $E(k)\sim k^{-5/3}$ \citep{M1982,sahraoui2009,andres2014,A2016b} to $E(k)\sim k^{-3/2}$ \citep{Scarf1967,matthaeus1989,bruno2005,podesta2007,Bhattacharjee2010}. The anisotropy of IMHD turbulence has been extensively studied in the literature
\citep{Robinson1971,Shebalin1983,Oughton1994,Bigot2008,D2021}. Consequently, several phenomenological theories drop the assumption of isotropy but in which the interactions between waves, and of waves with eddies, still play a central role \citep[e.g.,][]{Hig1984,Goldreich1995}. 

The understanding of the relation between waves and turbulence has been the subject of comprehensive research \citep{Fa2007,mininni2011,Me2013}. In order to identify the nature of waves in numerical simulations or experiments, the spatio-temporal spectra have been widely used  \citep{kobayashi2017,Me2016,Me2015,L2016,A2017a}. Using direct numerical simulations (DNSs) of the IMHD equations with a uniform magnetic field, \citet{Dmi2009} focused on the properties of fluctuations in the frequency domain. The authors found the presence of peaks at the corresponding Alfvén wave frequencies in the fully developed turbulent regimes, and non-linear transfer of energy at wave numbers perpendicular to the mean magnetic field. \citet{A2017a} studied the interaction between Alfvén waves and magnetosonic waves in a CMHD developed turbulent regime. Using spatio-temporal spectra, the authors showed direct evidence of excitation of Alfvén and magnetosonic waves at low Mach number. On the other hand, \citet{Me2015} performed three dimensional (3D) numerical simulations of incompressible weak MHD turbulence and found evidence of accumulation of energy in Alfvén waves and in intermittent structures, while \citet{Me2016} investigated the transition of turbulence from weak to strong regimes. Finally, \citet{L2016} considered relatively small, medium, and large values of the mean field field $B_0$ in IMHD turbulence. The authors found that time decorrelation of Fourier modes is dominated by sweeping effects, and only at large values of $B_0$ and for wave vectors mainly aligned with the guide field, time decorrelations are controlled by the Alfvénic time. 

The theory of compressible MHD turbulent flows has not been developed as much as incompressible ones, partially due to the fact that CMHD is more intricate because of non-linear coupling of velocity, magnetic field, density and pressure fluctuations \citep{Z1990,Z1992,Z1993}. This gives raise to two additional compressible modes (fast and slow magnetosonic waves as we mentioned earlier). These compressible modes can deeply affect the non-linear dynamics of turbulent plasmas. Moreover, these modes or their counterparts in the kinetic theory were reported using \textit{in situ} spacecraft measurements in the solar wind \citep{Kle2014,Wilson2016,ofman2016}, planetary magnetosheath \citep{S2003,S2006, romanelli2013,H2015,huang2015,huang2016}, and foreshock regions \citep{bertucci2007,andres2013,andres2015}. While fast modes can play an important role, for example in scattering of cosmic rays (note that the particle scattering critically depends on the properties of plasma turbulence), simulations have shown that the fast modes might only be a marginal component of compressible turbulence \citep{Yang2018}. However, these simulations have been driven incompressively by solenoidal forcing \citep{Vestuto_2003,Cho2003,Yang2018,Kowal2007,A2017a}. So a natural question to ask is whether and how the nature of forcing affects the interaction between waves in a turbulent regime. We try to answer this by compressively driving turbulence as we will see in the section \ref{sec:theory}. %no se si agregar lo de la seccion o si directamente mencionarlo al final de la introduccion.
Some earlier studies have driven turbulence by keeping a mixture of solenoidal and compressive velocity field at large scales \citep{YangYan2016} or by decomposing the driving force into solenoidal and compressive components \citep{Federrath2010}.  \citet{Makwana2020} studied the properties of MHD eigenmodes by decomposing the data of MHD simulations into linear MHD modes. The authors drove turbulence with a mixture of solenoidal and compressive driving and found that the proportion of fast and slow modes in the mode mixture increases with increasing compressive forcing. %citar lim?

Different theoretical and numerical efforts have been taken to understand the dynamics of compressible flows \citep{Z1993,Cho2002,Ga2011,B2013,Yang2017}. The nearly incompressible (NI) MHD theory is an intermediate model between compressible and incompressible descriptions. Using a particular expansion technique, \citet{Z1993} have derived different NI MHD equations depending on the $\beta$ plasma parameter (the ratio between fluid and magnetic pressure). From this NI perspective, one would expect that at high $\beta$ and low Mach number, the leading order description would be IMHD \citep{S2007}, with isotropic variances and anisotropic spectra. However, these theoretical predictions are subjected to initial conditions and forcing expressions. In contrast, the low $\beta$ NI MHD theory predicts an anisotropy in both the variances and the spatial spectra, which has been observed in solar wind \citep{Smith2006} and confirmed in several simulations (see e.g., \citet{Ou2016}). \citet{chandran2008} considered the low $\beta$ regime and derived a set of kinetic equations that provide an approximate description of non-linear processes in collisionless plasmas. Neglecting the slow magnetosonic branch, \citet{chandran2005} used this model to conclude that three-wave interactions transfer energy to high-frequency fast magnetosonic waves and to a lesser extent to high-frequency Alfvén waves. The author also predicted a $\sim k^{-7/2}$ power spectrum for the fast magnetosonic branch for low $\beta$ values.

The main objective of the present paper is to study the interaction between linear waves and turbulence by using 3D DNSs under the CMHD model. Therefore, we use the spatio-temporal spectra \citep{Cl2015}, which allow us to measure the amount of energy available in the different scales, both spatial and temporal, of the plasma. We keep in mind that in strong turbulence, much of the energy resides in modes that are not linear eigenmodes, but rather might be described as zero frequency turbulence. This analysis will be carried out for different types of kinetic forcing, ranging from a purely incompressible forcing to a purely compressible forcing. In addition, a parametric study will be performed varying both the Mach number and the amplitude of the mean magnetic field. The paper is organized as follows: in Sec.~\ref{sec:theory}, we present the CMHD model, and in Sec.~\ref{subsec:theory} and ~\ref{subsec:modes} we show the set of equations and normal modes of the CMHD model, respectively; in Sec.~\ref{subsec:NumericalSetUp} we describe the numerical set up used for the study and the equation for the forcing, introducing the parameter $f_0$ used to control the compressibility of the forcing; and in Sec.~\ref{subsec:SpatioTemporalSpectrum}, we briefly explain the spatio-temporal spectrum technique. In Sec. \ref{sec:results}, we present our results. Finally, in Sec.~\ref{sec:conclusion}, we summarize our main findings. 

%First, we study density fluctuations for different values of $f_0$, mean magnetic field and Mach number and then, we focus on the analysis of spatio-temporal spectrum of the magnetic field fluctuations and density fluctuations. Finally, we use an integral method in order to quantify the amount of energy located in each wave mode of the system. 

\section{Theory and numerical method} \label{sec:theory}

\subsection{Compressible MHD equations}\label{subsec:theory}
The 3D CMHD model is given by the mass continuity equation, the momentum equation, the induction equation for magnetic field and the polytropic state equation as,
\begin{equation}
    \frac{\partial \rho}{\partial t}+\boldsymbol\nabla\cdot(\rho \mathbf{u})=0,
    \label{continuidad}
\end{equation}
\begin{equation}
     \frac{\partial \mathbf{u}}{\partial t}+\mathbf{u} \cdot{\boldsymbol\nabla} {\bf u} =-\frac{{\boldsymbol\nabla} P}{\rho}+\frac{\mathbf{J}\times\mathbf{B}}{\rho}+\nu\bigg[\nabla^ 2 \mathbf{u}+\frac{{\boldsymbol\nabla} ({\boldsymbol\nabla}\cdot{\bf u})}{3}\bigg],
     \label{NS}
\end{equation}
\begin{equation}
    \frac{\partial \bf{B}}{\partial t}=\boldsymbol\nabla\times(\bf{u}\times \bf{B})+\eta \nabla^2 \bf{B},
    \label{induccion}
\end{equation}
\begin{equation}
    \frac{P}{\rho^\gamma}=\text{constant},
    \label{Politrópica}
\end{equation}
where $\bf{u}$ is the velocity field, $\bf{B}=\bf{B}_0+\bf{b}$ is the total magnetic field (with a fluctuating part $\bf{b}$ and a mean field $\bf{B}_0$). In addition, $\rho$ is the mass density, $P$ is the scalar isotropic pressure, $\bf{J}=\boldsymbol\nabla\times\bf{B}$ is the electric current, $\gamma=5/3$ is the polytropic index, and $\nu$ and $\eta$ are the kinematic viscosity and magnetic diffusivity, respectively. The main purpose of these last terms is to dissipate energy at scales smaller than MHD scales, while allowing us to study with an adequate scale separation compressible effects at the largest scales. In the present study, we take the viscosity and magnetic diffusivity to be independent of the mass density.

The set of equations \eqref{continuidad}-\eqref{Politrópica} can be written in a dimensionless form in terms of a characteristic length scale $L_0$, a mean scalar density $\rho_0$ and pressure $P_0$, and a typical magnetic and velocity field magnitude $b_{rms}$ and $u_{rms}=b_{rms}/\sqrt{4\pi\rho_0}$ (i.e., the r.m.s.~Alfvén velocity), respectively. Then, the unit time is $t_0=L_0/u_{rms}$, which for MHD becomes the Alfvén crossing time. The resulting dimensionless equations are,
\begin{equation}
    \frac{\partial \rho}{\partial t}+\boldsymbol\nabla\cdot(\rho \mathbf{u})=0,
    \label{continuidad_2}
\end{equation}
\begin{equation}
     \frac{\partial \mathbf{u}}{\partial t}+\mathbf{u} \cdot{\boldsymbol\nabla} {\bf u} =-\frac{1}{\gamma M_s^2}\frac{{\boldsymbol\nabla} P}{\rho}+\frac{\mathbf{J}\times\mathbf{B}}{\rho}+\nu'\bigg[\nabla^ 2 \mathbf{u}+\frac{{\boldsymbol\nabla} ({\boldsymbol\nabla}\cdot{\bf u})}{3}\bigg],
     \label{NS_2}
\end{equation}
\begin{equation}
    \frac{\partial \bf{B}}{\partial t}=\boldsymbol\nabla\times(\bf{u}\times \bf{B})+\eta' \nabla^2 \bf{B},
    \label{induccion_2}
\end{equation}
\begin{equation}
    \frac{P}{\rho^\gamma}=\text{constant},
    \label{Politrópica_2}
\end{equation}
where $M_s=u_{rms}/C_s$ is the Mach number, $C_s^2=\gamma P_0/\rho_0$ is the sound speed, and $\nu'$ and $\eta'$ are the dimensionless viscosity and magnetic diffusivity, respectively (written in terms of the number of the kinetic and magnetic Reynolds numbers). 

\subsection{Compressible MHD waves modes}\label{subsec:modes}

Considering a static equilibrium ($u_0=0$) with homogeneous external magnetic field ${\bf{B}_0}=B_0\hat{z}$, a constant density $\rho_0$, and a constant pressure $P_0$, we linearize Eqs.~\eqref{continuidad_2}-\eqref{Politrópica_2} and obtain the dispersion relation $\omega(\bf{k})$ of small amplitude waves propagation in a CMHD plasma. It is straightforward to show that there are
three independent propagating modes (or waves), which correspond to the so-called Alfvén waves (A), fast (F), and slow (S) magnetosonic waves \citep[e.g.,][]{fitzpatrick2014plasma},
\begin{align}\label{Alven_wave} 
    \omega^2_A(k)&=k_{\parallel}^2u_A^2, \\ \label{magnetosonic}
    \omega_{F,S}^2(k)&=k^2u_A^2\Bigg[\frac{(1+\beta)}{2}\pm \sqrt{\frac{{(1+\beta)}^2}{4}-\beta{\frac{k_{\parallel}}{k}}^2}\Bigg],
\end{align}
where $\beta={(C_s/u_A)}^2$ is the plasma beta, i.e., the ratio of plasma pressure to magnetic pressure; with $u_A=B_0/\sqrt{4\pi\rho_0}$ the Alfvén velocity, and $k=|{\bf k}|=\sqrt{{\bf k}_{\parallel}^2+{\bf k}_{\perp}^2}$ with ${\parallel}$ and $\perp$ the wave number component along and perpendicular to the external magnetic field, respectively. On one hand, Alfvén waves are incompressible fluctuations transverse to the external magnetic guide field ${\bf B}_0$. On the other hand, both fast and slow modes, unlike Alfvén modes, carry density fluctuations and their magnetic field perturbations have longitudinal and transverse components. In the case of fast modes, the magnetic field and the plasma are compressed by the wave motion, such that the restoring force is large and hence the frequency and the propagation speed are high. While, for slow modes, the magnetic field oscillation and the pressure variation are anti-correlated with each other such that the restoring force acting on the medium is weaker than that for the fast mode. For this reason the frequency and the propagation speed are the lowest among the three MHD waves modes. 
Note that for the perpendicular propagation (i.e., $k_{\parallel}=0$ and $k_{\perp}\ne 0$), the Alfvén and slow modes become non-propagating modes (i.e., $\omega_{A,S}=0$) and are degenerate, but they can be distinguished using their different polarization, since $\delta B_{\parallel A}=0$ and $\delta B_{\parallel S}\ne 0$. 

As one of the goals of the present paper is to identify the various possible waves and structures in the simulations, we adopt the assumption that energy concentrated closely to the linear dispersion relation can be explained by linear and weak turbulence theories, while any spread round, or away from, those linear curves is a sign of strong turbulence that requires fully non-linear theories to be understood \citep{A2017a}. 

\subsection{Numerical Setup}\label{subsec:NumericalSetUp}

We carried out direct numerical simulations (DNSs) of the compressible fluid equations in the presence of a mean magnetic field $\bf{B}_0$ in a periodic 3D box of size $2\pi$ with linear spatial resolutions of $128^3$ and $256^3$ grid points. The CMHD Eqs. \eqref{continuidad_2}-\eqref{Politrópica_2} were numerically solved using the Fourier pseudospectral code GHOST \citep{Go2005b,Mi2011}. The discrete time integration used is a second-order Runge–Kutta method. The scheme ensures exact energy conservation for the continuous time, spatially discrete equations \citep{Mi2011}.
For simplicity, we used identical dimensionless viscosity and magnetic diffusivity, $\nu'=\eta'=1.25\times10{-3}$ (i.e., the magnetic Prandtl number is $P_m=1$).

Random initial conditions were used for the velocity and magnetic field and null density fluctuations in the whole space. For all times $t>0$, the velocity field and the magnetic vector potential were forced by a mechanical forcing $F$ and electromotive forcing $\epsilon$, respectively. The mechanical and electromotive forcings are uncorrelated and they inject neither kinetic nor magnetic helicity. At $t=0$, for each forcing function, a random 3D isotropic field $f_k$ is generated in Fourier space, by filling the components of all modes in a spherical shell with $2<k<5$ with amplitude $f$, and a random phase $\phi_k$ for each wave vector $\bf{k}$. We used an amplitude $f=0.2$ for the mechanical and electromotive forcings. For the kinetic forcing, we introduced a parameter that allow us to determine its compressibility, so we can go from a pure incompressible forced system to a pure compressible forcing controlling this factor. The forcing is given by the equation \citep{J2018p},
\begin{equation}
    \bf{F}_k=i\bf{k}\times\hat{z}\psi_1-i\bf{k}\psi_2,
\end{equation}
where
\begin{equation}
    \psi_1=(1-f_0)f_1(\bf{k}),
\end{equation}
\begin{equation}
    \psi_2=f_0f_2(\bf{k}).
\end{equation}
Here $f_i(\bf{k})$ is a function that is $1$ if $\bf{k}$ is within the forced wave number range and is $0$ out of that range.
By modifying the value of the parameter $f_0$, the fluid can be compressibly excited ($f_0=1$) generating thrusts parallel to the direction of the movement and, alternatively, the incompressible modes are generated using $f_0=0$ pushing in the direction perpendicular to the fluid movement. By means of this forcing, we perfomed three groups of simulations: one with purely compressible forcing, another with purely compressible forcing and finally a mixed forcing that injects energy in the same amount to compressible and incompressible modes. In addition, for each of these three groups, four simulations were carried out for different values of the Mach number ($M_s$) and the mean magnetic field ($B_0$). We selected two different values of the sonic Mach number being $0.25$ and $0.55$, while for the magnetic field values, $B_0=2$ and $B_0=8$ were used. In Table \ref{table1}, we summarized the parameters of our simulations.

\subsection{Spatio-temporal spectrum}\label{subsec:SpatioTemporalSpectrum}

The spatio-temporal spectrum allows identification of waves in a turbulent flow and reveals aspects of its dynamics. The technique consists of calculating the complete spectrum in wave number and frequency for all available Fourier modes in a numerical simulation or an experiment \citep{Cl2015,Sahraoui2003}. As a result, it can separate between modes that satisfy a given dispersion relation (and are thus associated with waves) from those associated with nonlinear structures or turbulent eddies, and quantify the amount of energy carried by each of them at different spatial and temporal scales. The method we use does not require the pre-existence of wave modes or eddies. Quantifying the relative importance of each of them and understanding the physics that controls it is the main outcome expected from the present analysis. In the following, the spatio-temporal magnetic energy spectral density tensor is defined as:
\begin{equation}\label{ten_spec}
    E_{ij}({\bf k},\omega)=\frac{1}{2}\hat{{\bf B}}_i^*({\bf k},\omega)\hat{{\bf B}}_j({\bf k},\omega),
\end{equation}
where $\hat{\bf B}_i(\textbf{k},\omega)$ is the Fourier transform in space and time of the $i$-component of the magnetic field ${\bf B}({\bf x},t)$ and the asterisk implies the complex conjugate. The magnetic energy is associated with the trace of $E_{ij}(\textbf{k},\omega)$.

As the external magnetic field $\textbf{B}_0$ in the simulations
points in $\hat{z}$, in practice, we will consider either $i=j=y$ or $i=j=z$, to identify different waves based on their polarization (either transverse or longitudinal with respect to the guide field). In all cases, the acquisition frequency was at least two times larger than the frequency of the fastest wave, and the total time of acquisition was larger than the period of the slowest wave in the system. 
It is worth mentioning that spatio-temporal spectra have been used before in numerical simulations and experiments of rotating turbulence \citep{C2014}, stratified turbulence \citep{di2015}, quantum turbulence \citep{di2015b}, and IMHD turbulence simulations \citep{Me2015,Me2016,L2016} and in spacecraft observations \citep{Sahraoui2003,Sahraoui2010}. In the present paper, we use the technique to investigate the interaction between waves in CMHD turbulence extending our 
previous study \citep{A2017a}.

\begin{figure}[]
    \centering
    \subfigure{\includegraphics[width=0.3\linewidth]{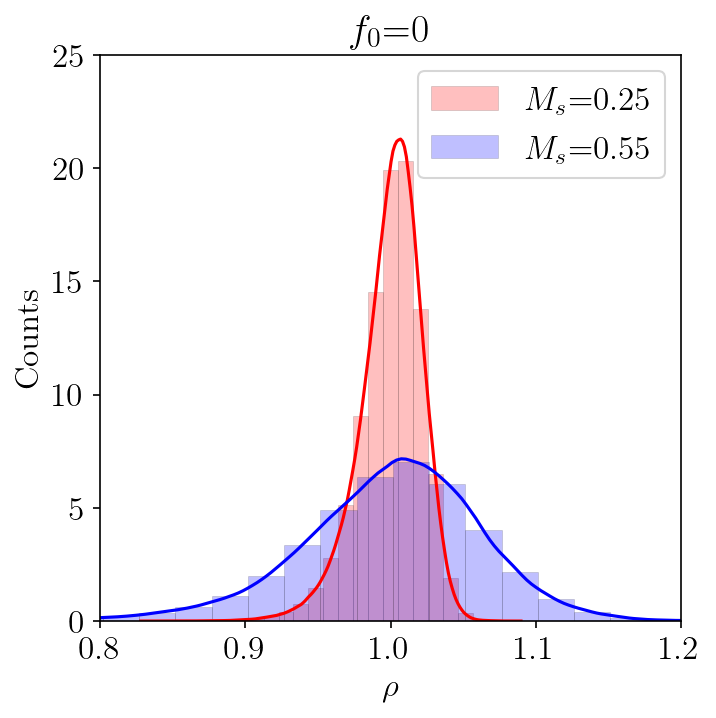}}
    \subfigure{\includegraphics[width=0.3\linewidth]{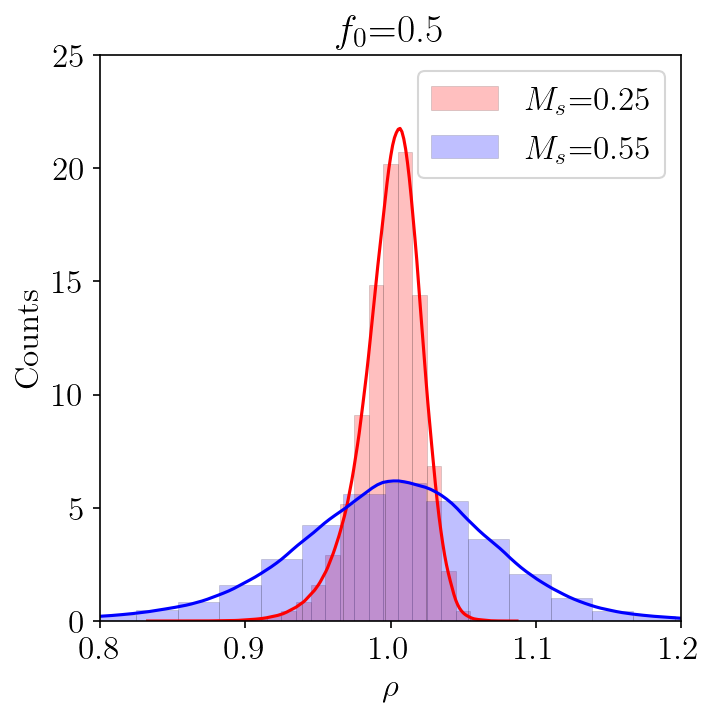}}
    \subfigure{\includegraphics[width=0.3\linewidth]{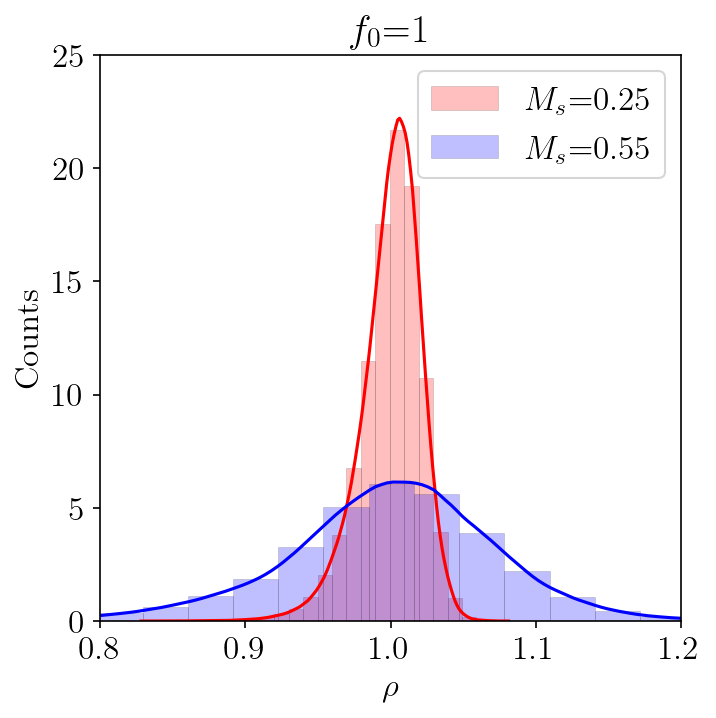}}
    \caption{Histogram of density fluctuations for fixed $B_0=0$ and taking different values of $M_s$ and $f_0$.}
    \label{Densidad_Hist}
\end{figure}

\begin{figure}[]
    \centering
    \subfigure{\includegraphics[width=0.3\linewidth]{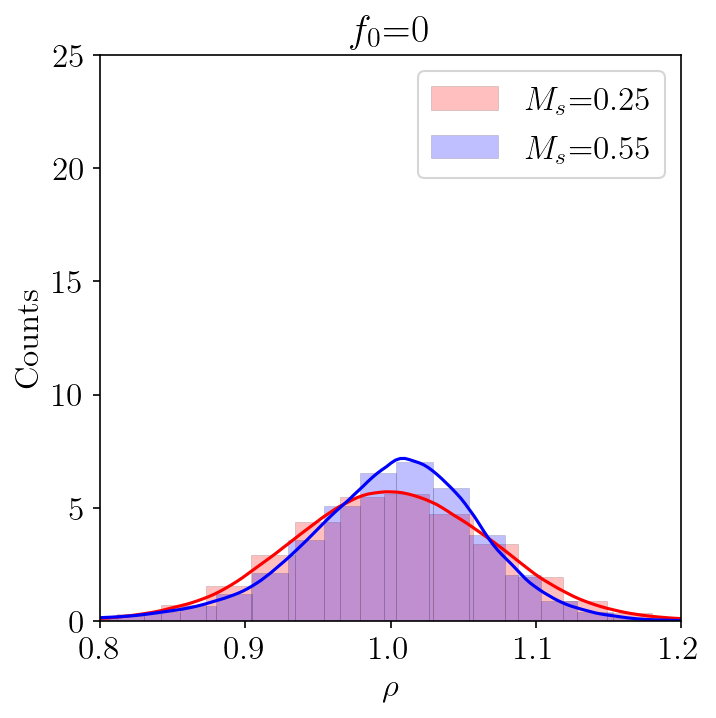}}
    \subfigure{\includegraphics[width=0.3\linewidth]{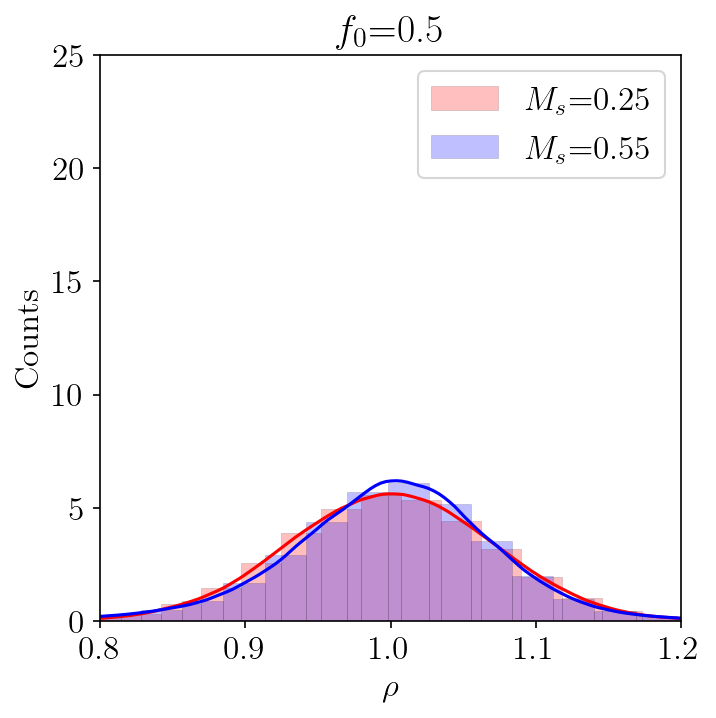}}
    \subfigure{\includegraphics[width=0.3\linewidth]{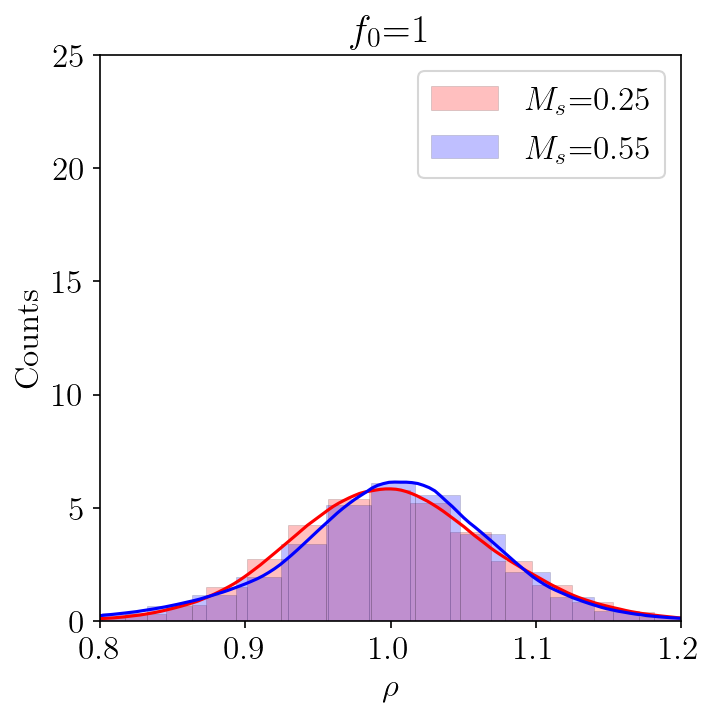}}
    \caption{Histogram of density fluctuations for fixed $B_0=2$ and taking different values of $M_s$ and $f_0$.}
    \label{Densidad_Hist_2}
\end{figure}

\begin{figure}[]
    \centering
    \subfigure{\includegraphics[width=0.3\linewidth]{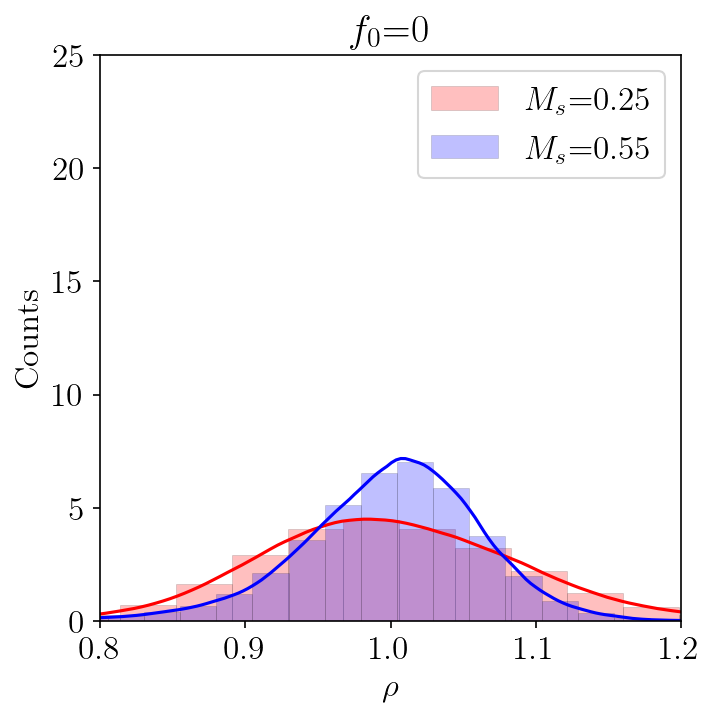}}
    \subfigure{\includegraphics[width=0.3\linewidth]{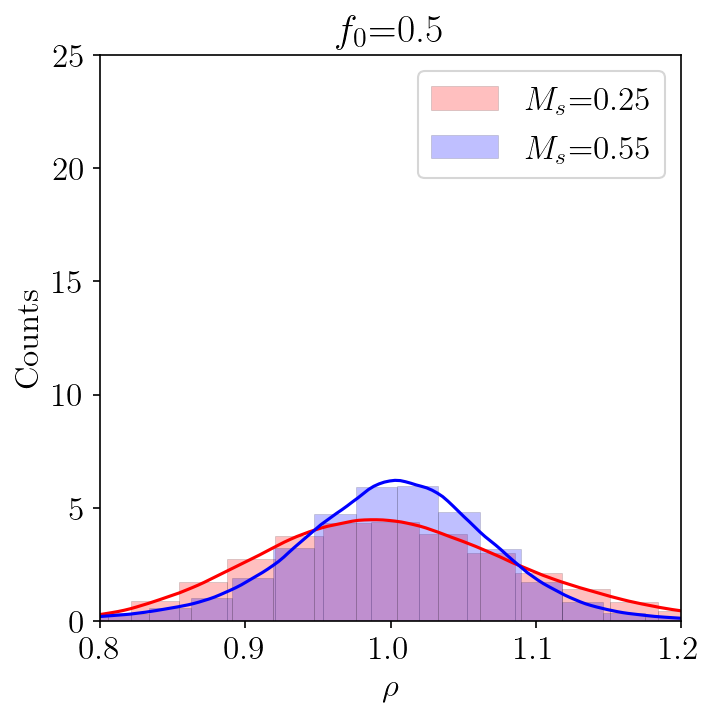}}
    \subfigure{\includegraphics[width=0.3\linewidth]{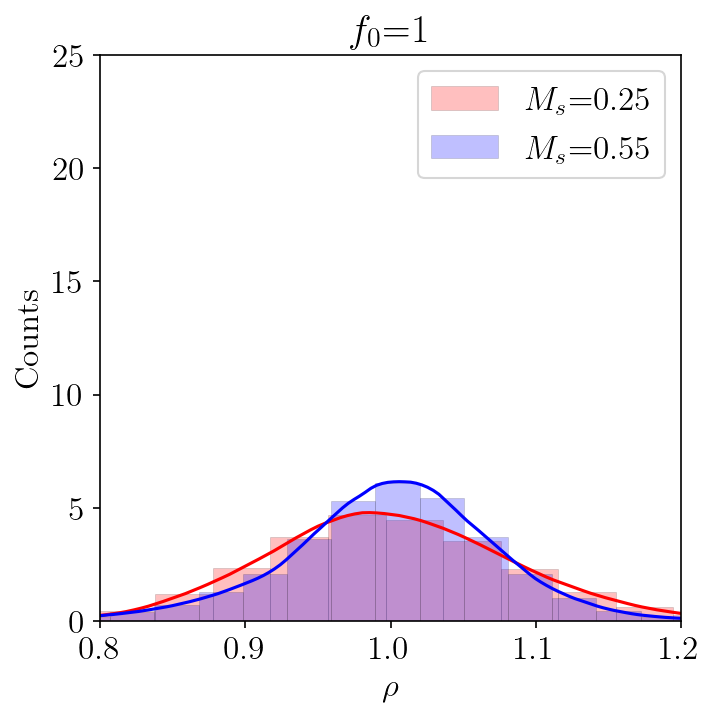}}
    \caption{Histogram of density fluctuations for fixed $B_0=8$ and taking different values of $M_s$ and $f_0$.}
    \label{Densidad_Hist_8}
\end{figure}

\section{Numerical Results} \label{sec:results}

We performed 3D DNS with small ($N=128$) and moderate ($N=256$) spatial resolution for different values of $f_0$, $M_s$ and $B_0$. The moderate resolution allows us to vary this extended parameter space and perform several simulations for extended periods of time (required for the spatio-temporal analysis). The small resolution DNSs were used in order to quickly study the parameter space of $f_0$, $M_s$ and $B_0$. We reported here only the results obtained using the moderate resolution runs (see Table \ref{table1} for details). 

\begin{table}[h]
\centering
\begin{tabular}{||c c c c c||}
 \hline 
 Run & $M_s$ & $B_0$ & $f_0$ & $N$ \\
 \hline\hline
 I1 & 0.25 & 2 & 0 & 128/256 \\ 
 I2 & 0.25 & 8 & 0 & 128/256 \\
 I3 & 0.55 & 2 & 0 & 128/256 \\
 I4 & 0.55 & 8 & 0 & 128/256 \\ [1ex]
 M1 & 0.25 & 2 & 0.5 & 128/256 \\ 
 M2 & 0.25 & 8 & 0.5 & 128/256 \\
 M3 & 0.55 & 2 & 0.5 & 128/256 \\
 M4 & 0.55 & 8 & 0.5 & 128/256 \\ [1ex]
 C1 & 0.25 & 2 & 1 & 128/256 \\ 
 C2 & 0.25 & 8 & 1 & 128/256 \\
 C3 & 0.55 & 2 & 1 & 128/256 \\
 C4 & 0.55 & 8 & 1 & 128/256 \\
 \hline
\end{tabular}
\caption{The values of $M_s$, $B_0$, $f_0$ and $N$ used in each simulation are specified. The notation I, M and C refer to the type of forcing (I:incompressible forcing, M:mixed forcing and C:compressible forcing). }
\label{table1}
\end{table}

\subsection{Density fluctuations}

Figures \ref{Densidad_Hist}, \ref{Densidad_Hist_2} and \ref{Densidad_Hist_8} show histograms of density fluctuations for different Mach numbers $M_s$, mean magnetic fields $B_0$ and values of the compressibility amplitude of the forcing $f_0$. For the present analysis, we use some statistics tools (such as the variance and quantiles) in order to improve the description of density variations. In the case of $B_0=0$, numerical results indicate an increase in density fluctuations as the Mach number increases. For example, we find that for fixed $f_0=0$ and $M_s=0.25$, the $90\%$ of the data is located around the interval $\rho=[0.96-1.03]$ while for $M_s=0.55$ this happens around $\rho=[0.89-1.09]$. For $f_0=0.5$ and $f_0=1$, results are analogous. In particular, for $M_s=0.25$ in the vast majority of cases, the density does not vary being the simulation that most closely resembles to an incompressible plasma. The same occurs when the compressibility amplitude of the forcing is increased, and although the differences are less noticeable, we observe that for $M_s=0.55$ there is a slight increase in the density fluctuations with a variance ranging from $3.8\times10^{-3}$ ($f_0=0$) to $5.1\times10^{-3}$ ($f_0=1$).

Turning on the mean magnetic field $B_0$, we can notice a significant increase in the variance of about an order of magnitude, ranging from $4.5\times10^{-4}$ ($B_0=0$) to $4.8\times10^{-3}$ ($B_0=2$) for fixed $f_0=0$ and $M_s=0.25$. When we increase $B_0$, the density fluctuations grows with a variance of $4.8\times10^{-3}$ ($B_0=2$) and $7.8\times10^{-3}$ ($B_0=8$). Therefore, the $90\%$ of the data is located around the intervals $\rho=[0.88-1.11]$ ($B_0=2$) and $\rho=[0.86-1.15]$ ($B_0=8$). However, for different values of $M_s$ and $f_0$ and fixed $B_0$, the growth of density fluctuations is more subtle. For instance, in the case of $B_0=2$ and $M_s=0.55$, we have an increase of fluctuations as we vary $f_0$ with a variance of $3.8\times10^{-3}$ ($f_0=0$), $4.9\times10^{-3}$ ($f_0=0.5$) and $5.1\times10^{-3}$ ($f_0=1$). For $B_0=8$ the results are similar as shown in Figure \ref{Densidad_Hist_8}. Therefore, this analysis allows us to show that we have larger fluctuations as we increase sonic Mach number (approaching to $M_s=1$ and therefore, at speeds comparable to the speed of sound), the compressibility amplitude of the forcing and the mean magnetic field.

\begin{figure}[]
     \centering
     \begin{subfigure}
         \centering
         \includegraphics[width=1.0\linewidth]{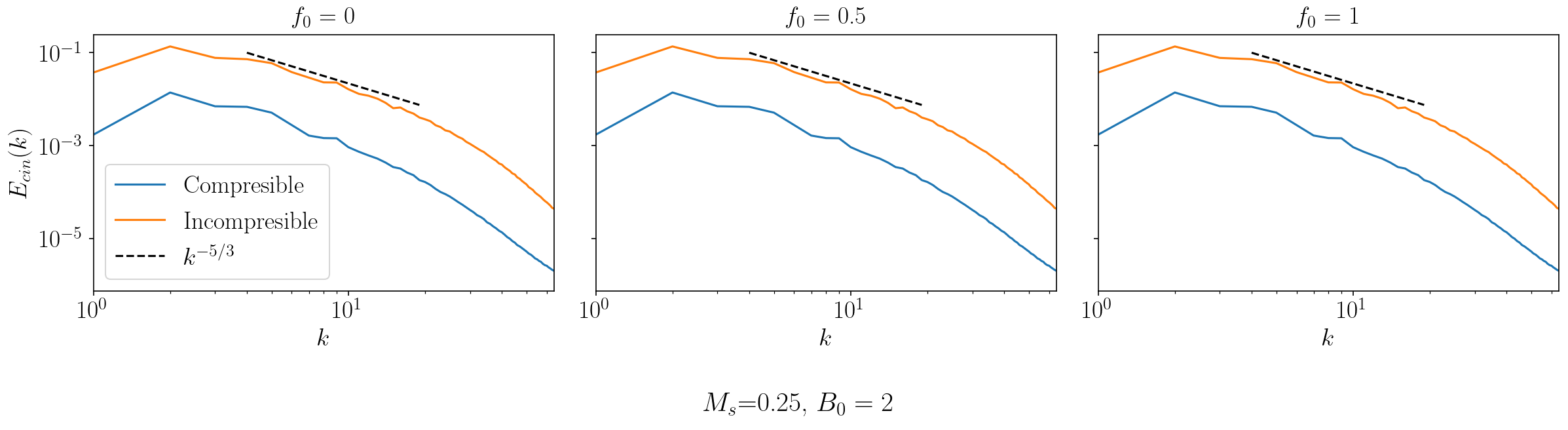}
         \label{espectro_b0_2_ms_025}
     \end{subfigure}
     \hfill
     \begin{subfigure}
         \centering
         \includegraphics[width=1.0\linewidth]{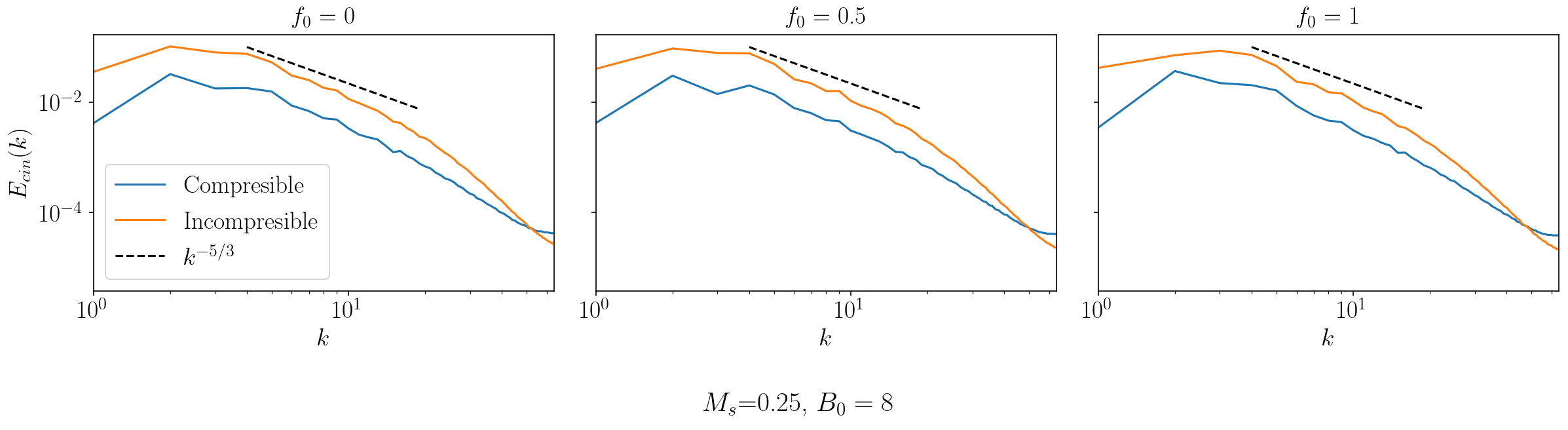}
         \label{espectro_b0_8_ms_025}
     \end{subfigure}
     \hfill
     \begin{subfigure}
         \centering
         \includegraphics[width=1.0\linewidth]{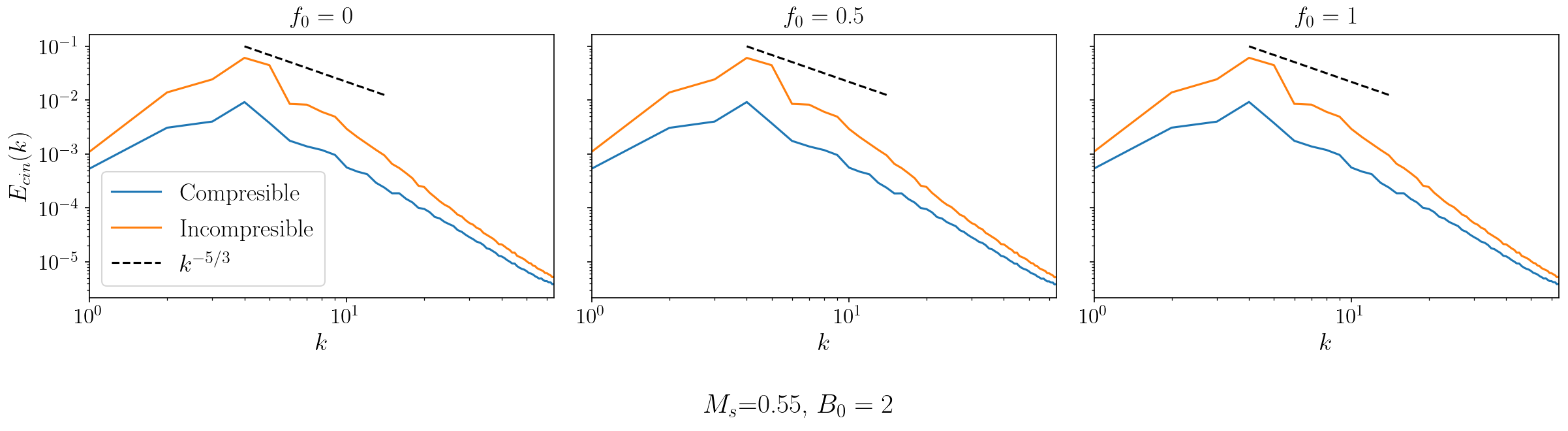}
         \label{espectro_b0_2_ms_055}
     \end{subfigure}
     \hfill
     \begin{subfigure}
         \centering
         \includegraphics[width=1.0\linewidth]{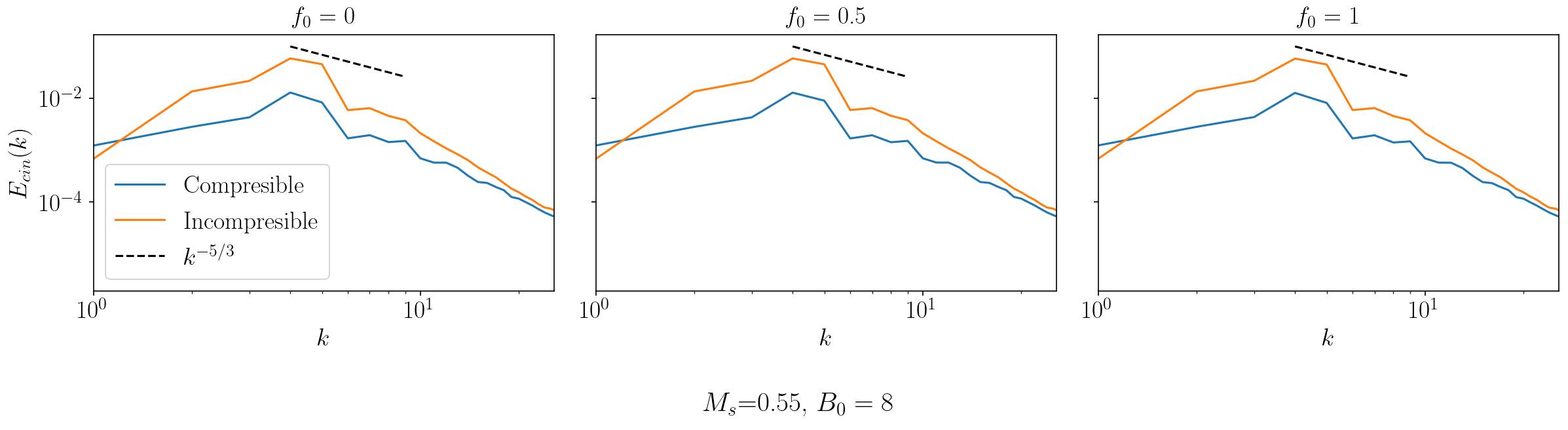}
         \label{espectro_b0_8_ms_055}
     \end{subfigure}
    \caption{Spatial spectrum of the kinetic energy for the set of simulations I1/M1/C1, I2/M2/C2, I3/C3/M3 and I4/M4/C4. The dotted line indicates the inertial range characterized by Kolmogorov's scaling law}
    \label{espectros_espaciales}
        \label{fig:fig}
\end{figure}

\subsection{Spatial spectrum}

In order to study the types of fluctuations present in a spatial spectrum, we computed the spatial isotropic spectrum for the kinetic energy. Besides, we computed the compressible and incompressible kinetic spectra of the flow using the usual Helmholtz decomposition \citep{J2006}. Figure \ref{espectros_espaciales} shows the spatial spectrum for $M_s=0.25,0.55$, $B_0=2,8$ and $f_0=0,0.5,1$ respectively. Also, for comparative purpose, in dotted line we include the scaling law $k^{-5/3}$, predicted by Kolmogorov theory for incompressible, isotropic and homogeneous turbulent flows. On the one hand, in the case $M_s=0.25$ and $B_0=2,8$ for a large range of wave numbers, an inertial range compatible with a $\sim k^{-5/3}$ can be observed for the incompressible and compressible kinetic energy. Whereas, when we take $M_s=0.55$ and same values of $B_0$, the compatibility persists for a narrower range of wave numbers. This means that we find important differences in the shape of the spectrum when increasing the Mach number, while the increase in the magnetic field does not imply significant changes. Note also that, although the vast majority of the kinetic energy is located in its incompressible component and dominates over the compressible one, it is known that the small compressible component can still affect the flow dynamics in this regime. For instance, DNSs showed that proton acceleration is significantly enhanced when compared to the incompressible case \citep{G2016}. Also, recently it has been showed that small compressibility in the plasma increases the energy dissipation rate as the cascade enters into the sub-ion scales \citep{A2019}, affects the anisotropic nature of solar wind magnetic turbulence fluctuations \citep{K2013}, and increases the turbulent energy cascade rate in the inner heliosphere \citep{A2021}.

\begin{figure}[]
    \centering
    \includegraphics[width=1.0\linewidth]{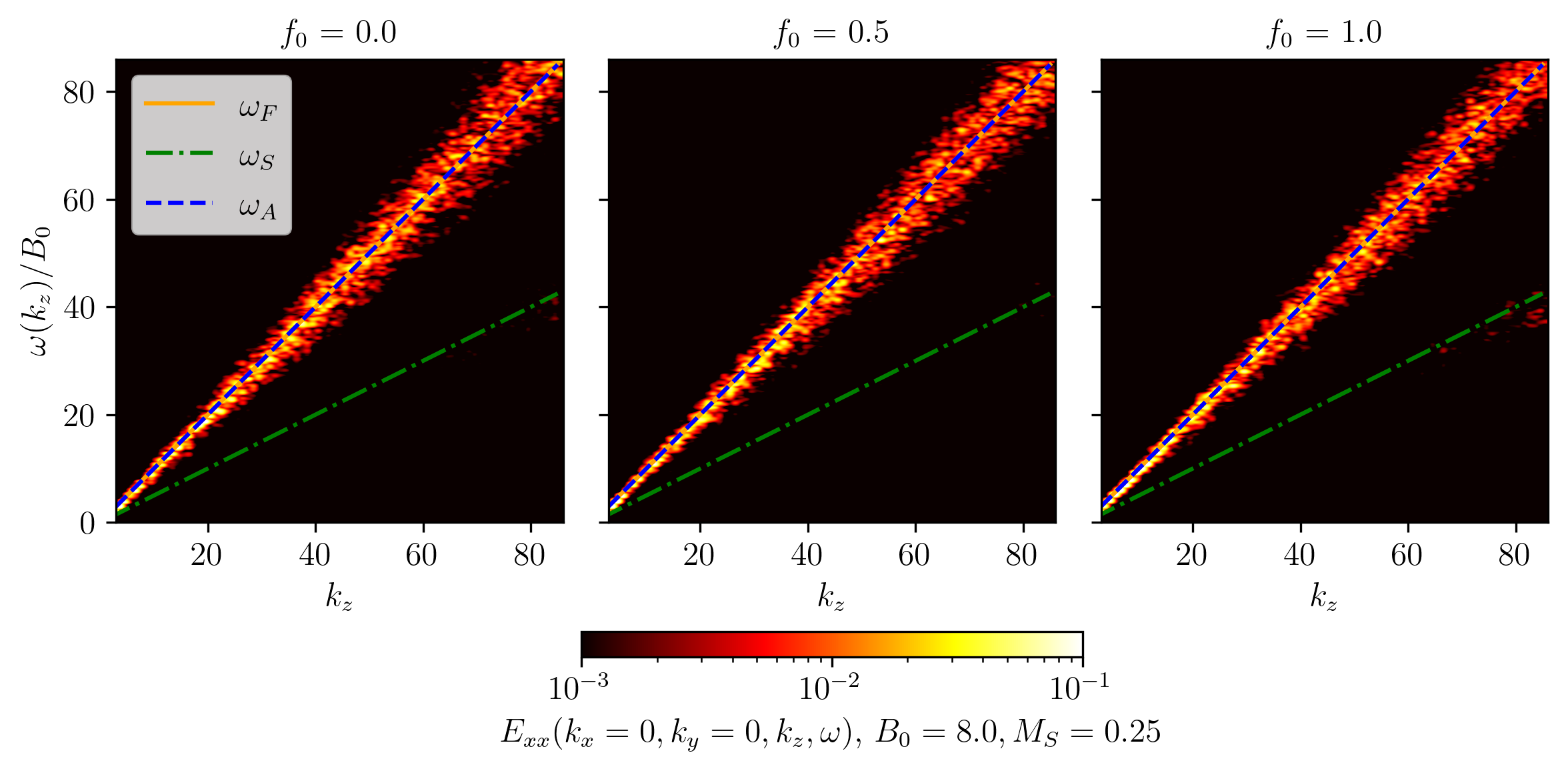}
    \caption{Spatio-temporal spectrum $E_{xx}(k_x=0,k_y=0,k_z)$ for the magnetic field fluctuations perpendicular to $B_0$, for the runs I2/M2/C2 taking different values of the compressibility amplitude of the forcing $f_0$. The spectrum is shown as a function of $\omega$ and $k_{\parallel}$ for fixed $k_x=k_y=0$. The blue dashed, green dashed-dotted and orange solid lines correspond to the linear dispersion relation of Alfvén waves ($\omega_A$), of fast magnetosonic waves ($\omega_F$) and of slow magnetosonic waves ($\omega_S$), respectively.}
    \label{ekw_bx_025}
\end{figure}

\begin{figure}[]
    \centering
    \includegraphics[width=1.0\linewidth]{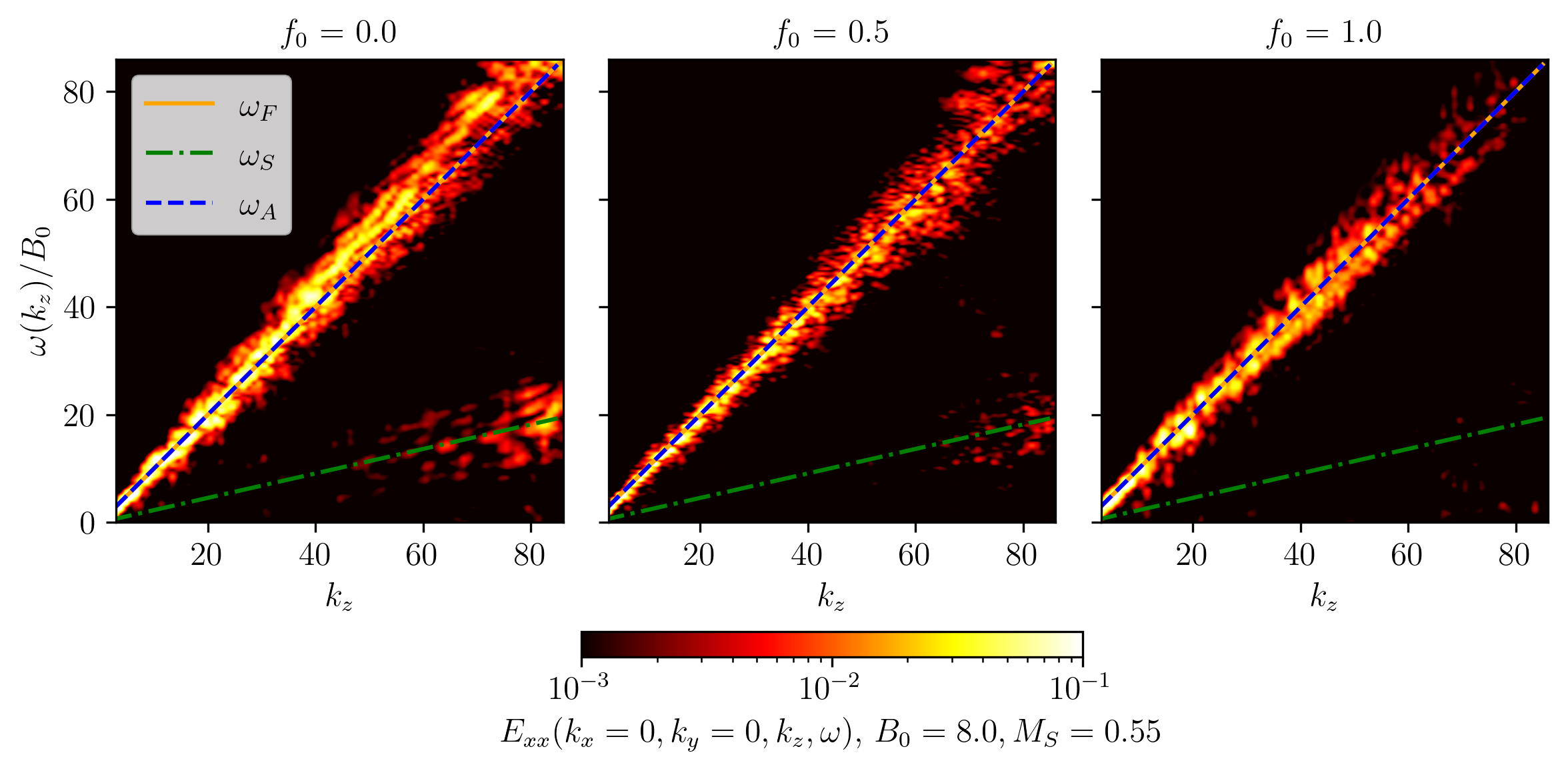}
    \caption{Spatio-temporal spectrum $E_{xx}(k_x=0,k_y=0,k_z)$ for the magnetic field fluctuations perpendicular to $B_0$, for the runs I4/M4/C4 taking different values of the compressibility amplitude of the forcing $f_0$. The spectrum is shown as a function of $\omega$ and $k_{\parallel}$ for fixed $k_x=k_y=0$. The blue dashed, green dashed-dotted and orange solid lines correspond to the linear dispersion relation of Alfvén waves ($\omega_A$), of fast magnetosonic waves ($\omega_F$) and of slow magnetosonic waves ($\omega_S$), respectively.}
    \label{ekw_bx_055}
\end{figure}

\begin{figure}[]
    \centering
    \includegraphics[width=1.0\linewidth]{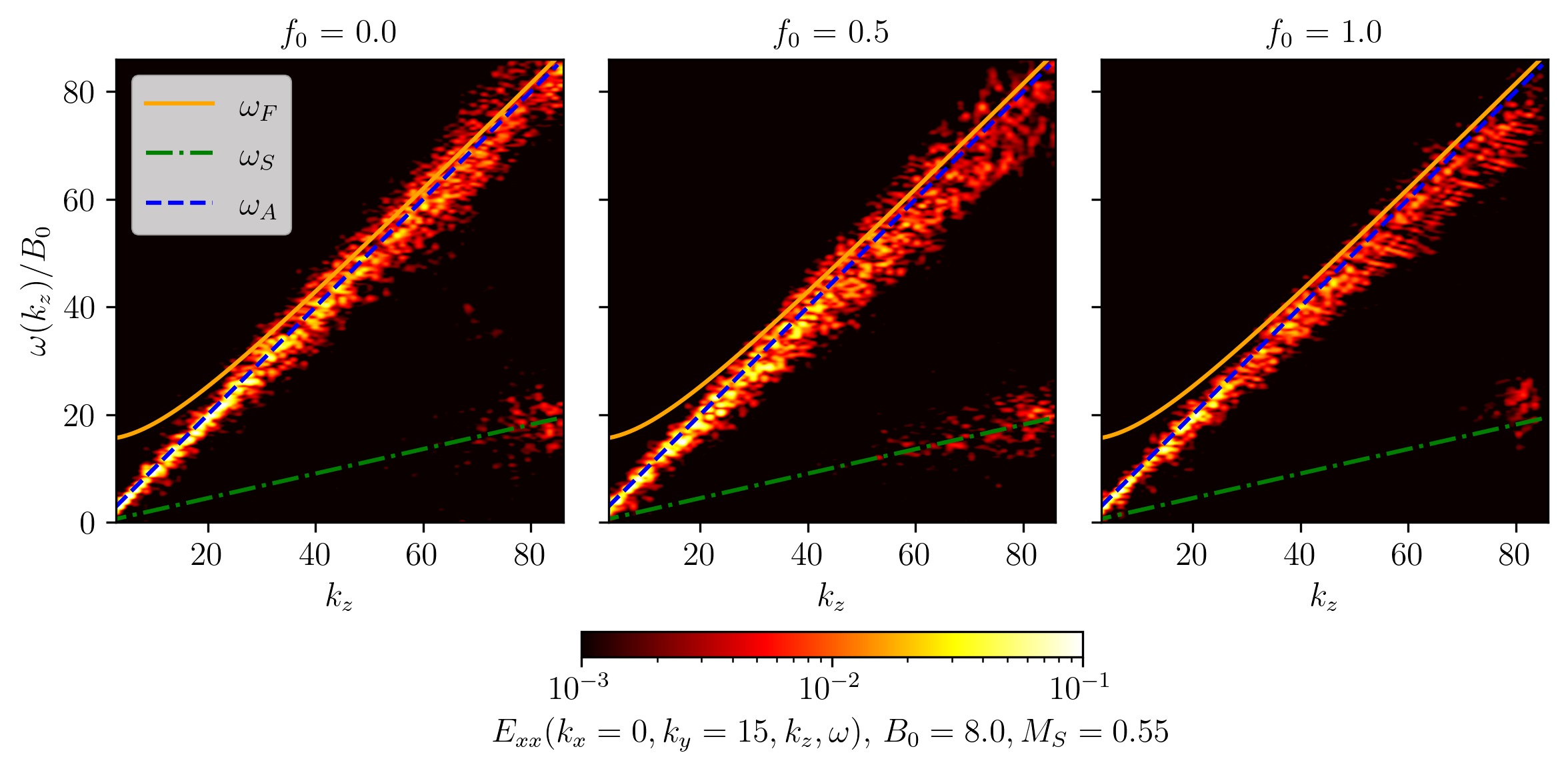}
    \caption{Spatio-temporal spectrum $E_{xx}(k_x=0,k_y=15,k_z)$ for the magnetic field fluctuations perpendicular to $B_0$, for the runs I4/M4/C4 taking different values of the compressibility amplitude of the forcing $f_0$. The spectrum is shown as a function of $\omega$ and $k_{\parallel}$ for fixed $k_x=0$ and $k_y=15$. The blue dashed, green dashed-dotted and orange solid lines correspond to the linear dispersion relation of Alfvén waves ($\omega_A$), of fast magnetosonic waves ($\omega_F$) and of slow magnetosonic waves ($\omega_S$), respectively.}
    \label{ekw_bx_055_ky_15}
\end{figure}

\begin{figure}[]
    \centering
    \includegraphics[width=1.0\linewidth]{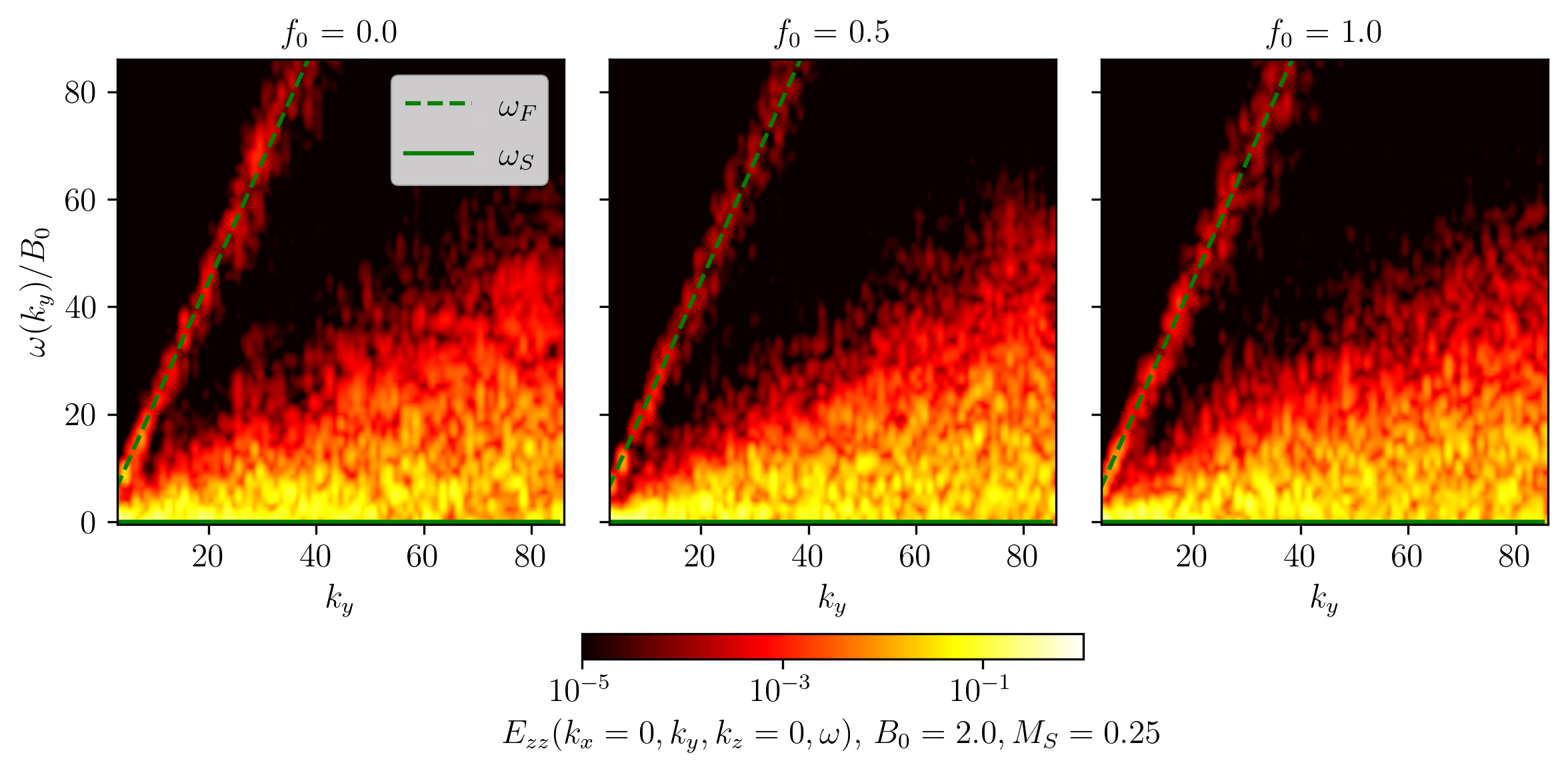}
    \caption{Spatio-temporal spectrum $E_{zz}(k_x=0,k_y,k_z=0)$ for the magnetic field fluctuations parallel to $B_0$, for the runs I1/M1/C1 taking different values of the compressibility amplitude of the forcing $f_0$. The spectrum is shown as a function of $\omega$ and $k_{y}$ for fixed $k_x=k_{\parallel}=0$. The dashed and the solid lines correspond to the linear dispersion relation of fast magnetosonic waves ($\omega_F$) and of slow magnetosonic waves ($\omega_S$), respectively.}
    \label{ekwz_bx_025_b0_2}
\end{figure}

\begin{figure}[]
    \centering
    \includegraphics[width=1.0\linewidth]{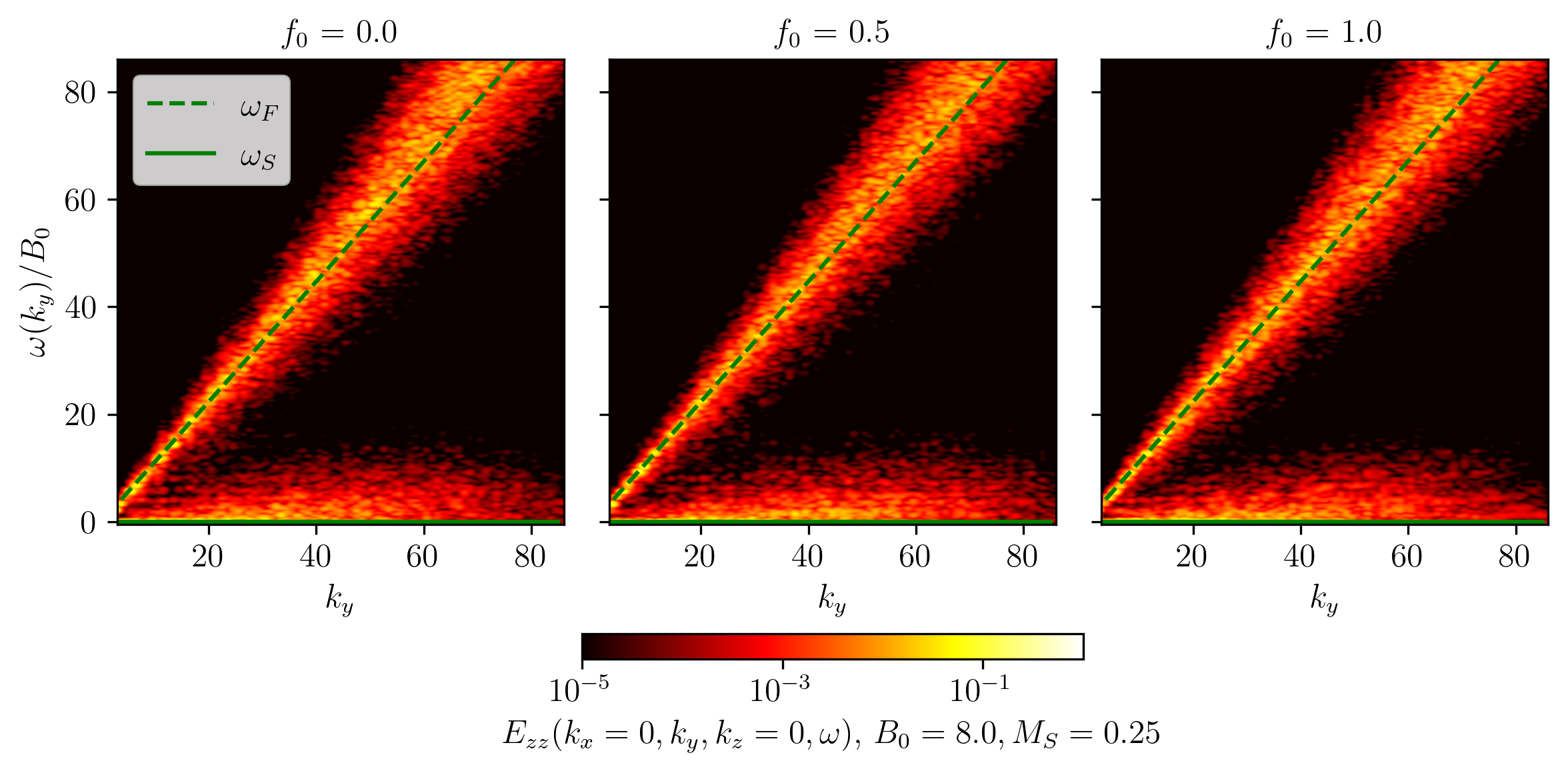}
    \caption{Spatio-temporal spectrum $E_{zz}(k_x=0,k_y,k_z=0)$ for the magnetic field fluctuations parallel to $B_0$, for the runs I2/M2/C2 taking different values of the compressibility amplitude of the forcing $f_0$. The spectrum is shown as a function of $\omega$ and $k_{y}$ for fixed $k_x=k_{\parallel}=0$. The dashed and the solid lines correspond to the linear dispersion relation of fast magnetosonic waves ($\omega_F$) and of slow magnetosonic waves ($\omega_S$), respectively.}
    \label{ekwz_bx_025_b0_8}
\end{figure}

\subsection{Spatio-temporal spectrum}

To determine the presence of Alfvén or magnetosonic waves and which (if any) dominates the dynamics, we use a spatio-temporal analysis \citep[see,][]{Cl2015}. To perform this type of analysis, the field must be stored with very high frequency cadence in order to resolve the waves in time and space; in particular $dt=1\times10^{-5}$ was taken as the temporal sampling rate. Since the spatio-temporal spectrum is four dimensional (with $\textbf{k}=(k_x,k_y,k_z)$ and the frequency $\omega$), we fix two components of $\textbf{k}$ to plot the remaining component against the frequency.

Figures \ref{ekw_bx_025} and \ref{ekw_bx_055} show the spatio-temporal spectrum of the perpendicular magnetic field fluctuation $E_{xx}(k_x=0,k_y=0,k_z)$ (with $k_z=k_{\parallel}$) for fixed $k_x=k_y=0$ for the set of simulations I2/M2/C2 and I4/M4/C4, respectively. The dispersion relation for Alfvén and magnetosonic waves given by Eqs. \eqref{Alven_wave} and \eqref{magnetosonic} are shown in blue dashed, green dashed-dotted and orange solid lines, respectively. In the case of runs I2/M2/C2, the wave modes that are excited correspond to fast magnetosonic and Alfvén branches (note that for $k_{\perp}=0$, these branches overlap) and there is no apparent traces of slow magnetosonic waves. Instead, for runs I4/M4/C4, the energy accumulates mainly in Alfvén and fast branches with a small portion of energy spread along the slow magnetosonic modes at high parallel wave numbers and decreasing as the compressibility amplitude of the forcing grows. In addition, we can observe that when $f_0=1$ at high wave numbers, the energy reaches its minimum with a maximum in $k_{\parallel}<25$ and $\omega/B_0<20$. It is worth noting that, as we increase the Mach number, the energy near the Alfvén/fast branches also increase by approximately an order of magnitude. This may be due to the contribution of the compressible modes given by fast magnetosonic waves. 

\begin{figure}[]
    \centering
    \subfigure{\includegraphics[width=0.4\linewidth]{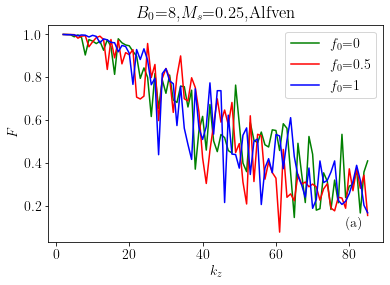}}
    \subfigure{\includegraphics[width=0.4\linewidth]{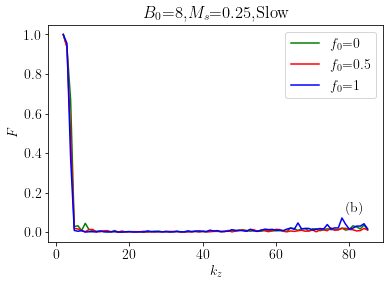}}
    \caption{Quantification of the amount of energy in runs I2/M2/C2 around Alfvén and slow magnetosonic branches present in spatio-temporal spectrum $E_{xx}(k_x=0,k_y=0,k_z)$.}
    \label{cuanti_b0_8_bx_025}
\end{figure}

\begin{figure}[]
    \centering
    \subfigure{\includegraphics[width=0.4\linewidth]{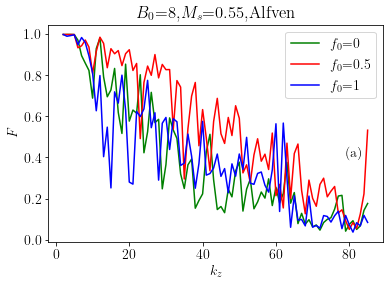}}
    \subfigure{\includegraphics[width=0.4\linewidth]{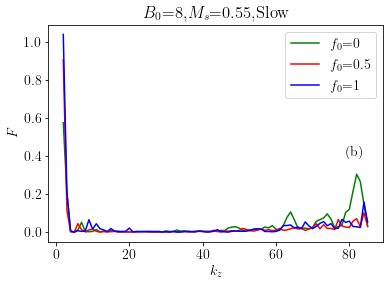}}
    \caption{Quantification of the amount of energy in runs I4/M4/C4 around Alfvén and slow magnetosonic branches present in spatio-temporal spectrum $E_{xx}(k_x=0,k_y=0,k_z)$.}
    \label{cuanti_b0_8_bx_055}
\end{figure}

\begin{figure}[]
    \centering
    \subfigure{\includegraphics[width=0.4\linewidth]{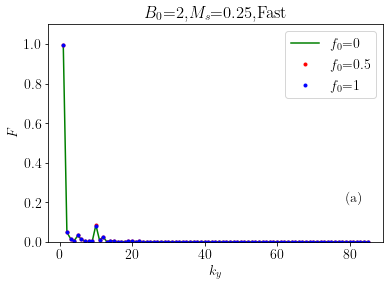}}
    \subfigure{\includegraphics[width=0.4\linewidth]{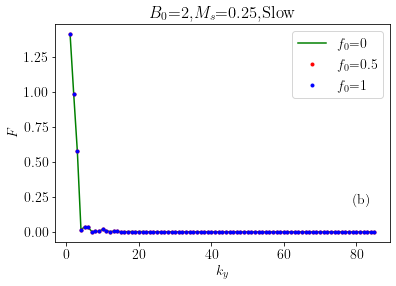}}
    \caption{Quantification of the amount of energy in runs I1/M1/C1 around Alfvén and slow magnetosonic branches present in spatio-temporal spectrum $E_{zz}(k_x=0,k_y,k_z=0)$.}
    \label{cuanti_b0_2_bz_025}
\end{figure}

\begin{figure}[]
    \centering
    \subfigure{\includegraphics[width=0.4\linewidth]{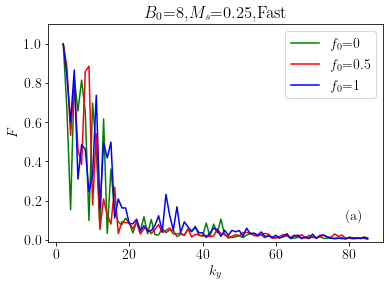}}
    \subfigure{\includegraphics[width=0.4\linewidth]{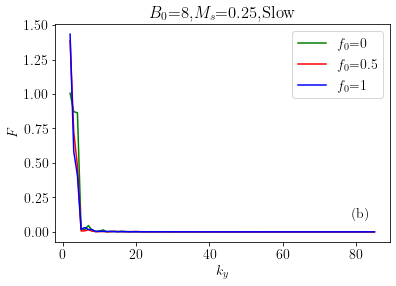}}
    \caption{Quantification of the amount of energy in runs I2/M2/C2 around Alfvén and slow magnetosonic branches present in spatio-temporal spectrum $E_{zz}(k_x=0,k_y,k_z=0)$.}
    \label{cuanti_b0_8_bz_025}
\end{figure}

From these last two spectra, the question of which branch dominates the dynamics of the system may arise: is it Alfvén or fast magnetosonic waves? Looking back to spatial spectra and the fact that the incompressible component dominates over the compressible one, we came to the conclusion that in both spatio-temporal spectrum, energy accumulates mainly around Alfvén modes. For spatio-temporal spectrum $E_{xx}(k_x=0,k_y=15,k_z)$ for fixed $k_y=15$, shown in Figure \ref{ekw_bx_055_ky_15}, a great amount of energy accumulates around the Alfvén wave branch. Although some energy is also present in the vicinity of the fast magnetosonic branch and along slow modes at high parallel wave numbers, fast waves do not dominate the dynamics as predicted using the weak wave turbulence theory \citep{chandran2005,chandran2008}. Instead, Alfvén waves dominate the linear dynamics as we suspected from the spatial spectra results. Besides, lower magnetic energy compared to the case $k_y=0$ is observed for the different values of $f_0$. This is because, when we choose $k_y=15$, we are moving along the inertial range and, according to previous results \citep{A2017a}, the presence of waves tends to decrease.

Fast magnetosonic waves can be separated from the Alfvén waves by looking at the spatio-temporal spectrum of parallel magnetic field fluctuation $E_{zz}(k_x=0,k_y,k_z=0$) shown in Figures \ref{ekwz_bx_025_b0_2} and \ref{ekwz_bx_025_b0_8} for the set of simulations I1/M1/C1 and I2/M2/C2. The Alfvén waves do not contribute to the parallel component of the magnetic field energy since their magnetic perturbations are perpendicular to the guide field. In both figures, we find that the energy accumulates in two regions: at high frequencies near the fast magnetosonic branch and at low frequencies near slow modes with $\omega/B_0=0$. We can observe that as we increase the mean magnetic field, we go from having most of the energy spread across the spectrum to being located around the fast and slow modes. A possible explanation is that, when we increase the magnetic field $B_0$, the anisotropic turbulence grows, the energy is concentrated on non-propagating modes (with $\omega=0$) and produce a cascade in the wave vectors perpendicular to the mean magnetic field direction only \citep{Matthaeus1996,Oughton1998,O2015}. 

\begin{figure}[]
    \centering
    \includegraphics[width=1.0\linewidth]{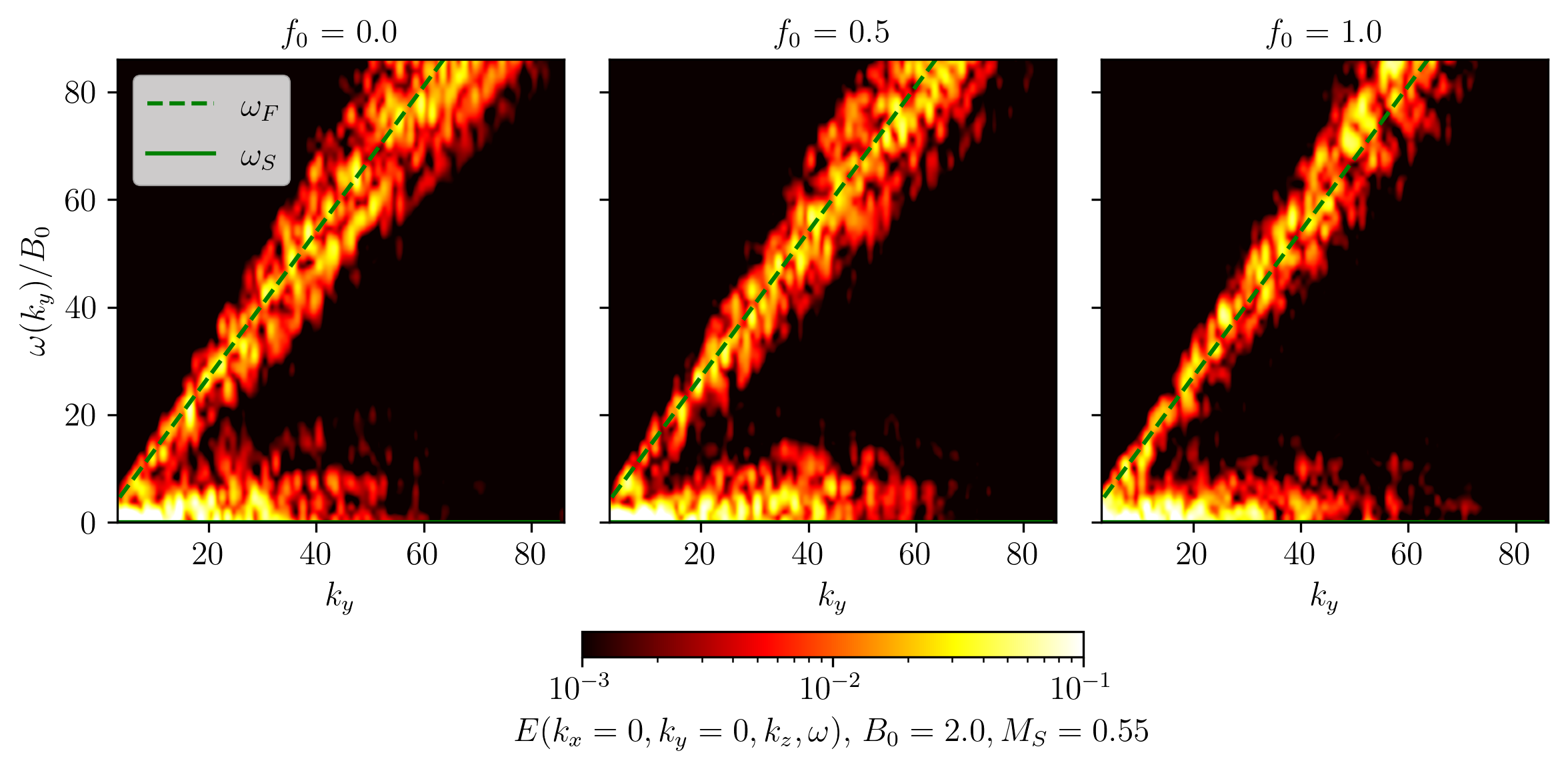}
    \caption{Spatio-temporal spectrum $E_(k_x=0,k_y,k_z=0)$ for the density fluctuations, for the runs I3/M3/C3 taking different values of the compressibility amplitude of the forcing $f_0$. The spectrum is shown as a function of $\omega$ and $k_{y}$ for fixed $k_x=k_{\parallel}=0$. The dashed and the solid lines correspond to the linear dispersion relation of fast magnetosonic waves ($\omega_F$) and of slow magnetosonic waves ($\omega_S$), respectively.}
    \label{ekwz_th_055_b0_2}
\end{figure}

\begin{figure}[h]
    \centering
    \subfigure{\includegraphics[width=0.4\linewidth]{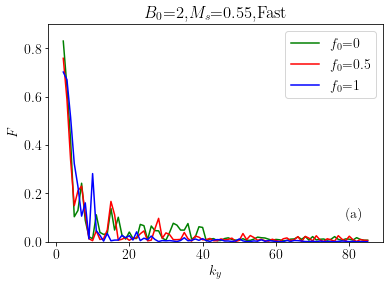}}
    \subfigure{\includegraphics[width=0.4\linewidth]{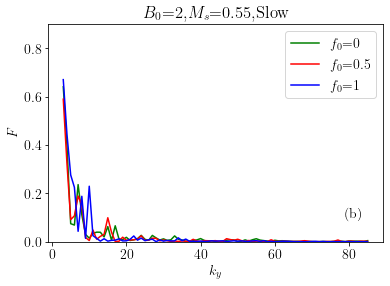}}
    \subfigure{\includegraphics[width=0.4\linewidth]{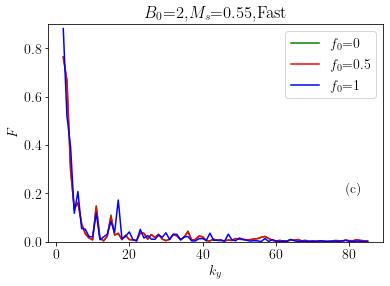}}
    \subfigure{\includegraphics[width=0.4\linewidth]{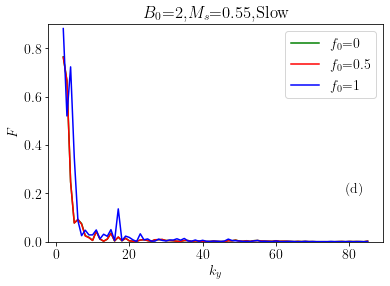}}
    \caption{Quantification of the amount of energy in runs I3/M3/C3 around fast and slow magnetosonic branches present in spatio-temporal spectrum $E(k_x=0,k_y,k_z=0)$ for density fluctuations ((a) and (b)) and magnetic field fluctuations ((c) and (d)).}
    \label{cuanti_b0_th_2_bx_055}
\end{figure}

In order to quantify the amount of energy near the different wave modes, we use an integration method available in the literature. \citet{Cl2015} propose to calculate the ratio of the energy accumulated near these modes to the total energy in the same wave number,
\begin{equation}
    F(k_z)=\frac{E_{xx}(k_x=0,k_y=0,k_z,\omega=\omega_{a,f,s})}{\Sigma_j E_{xx}(k_x=0,k_y=0,k_z,\omega_j)},
    \label{cuantificacion}
\end{equation}
with $\omega_{a,f,s}$ the frequencies that satisfy a certain dispersion relation (we have used the $E_{xx}(k_z)$ component as an illustration example only). Figure \ref{cuanti_b0_8_bx_025} shows the energy near Alfvén and slow magnetosonic waves for runs I2/M2/C2. We can observe that energy is located mainly in the Alfvén branch with a maximum in the smallest wave numbers. Analysing runs I4/M4/C4 (shown in Figure  \ref{cuanti_b0_8_bx_055}), for $f_0=0$ and large scales of $k_{\parallel}$, we obtain an increase of energy around slow modes. However, the vast majority of the energy is located around the Alfvén branch with a maximum in $k_{\parallel}<60$ and $f_0=0.5$. Figure \ref{cuanti_b0_2_bz_025} shows that, for runs I1/M1/C1, the amount of energy located both in the fast branch and in the slow one is similar. Whereas, for runs I2/M2/C2 (Figure \ref{cuanti_b0_8_bz_025}), there is a greater amount of energy around the fast magnetosonic branch with a maximum in $k_{\parallel}<20$. In all cases, as we increase the value of $f_0$, no significant growth of energy was found in the different branches. It is worth mentioning that we have also used an alternative criterion to compute the amount of energy in each wave branch reported by \citet{J2018p} and \citet{J2018Thesis}. This method consists of performing the mean squared differences between the frequencies that follow a certain dispersion relation $\omega_{a,f,s}$ and the spectrum, e.g., $E_{xx}(k_x=0,k_y=0,k_z)$. The results (not shown here) are similar to the ones present here and, therefore both methods are compatible.

Finally, we investigate the spatio-temporal spectrum of density defined as, $E(k,\omega)\sim\delta \rho^{*}\delta \rho$, where $\delta\rho=\rho-\rho_0$ represents the density fluctuations. Figure \ref{ekwz_th_055_b0_2} shows the spatio-temporal spectrum $E(k_x=0,k_y,k_{\parallel}=0)$ of density fluctuations for runs I3/M3/C3. We observe that, energy is spread along both fast and slow magnetosonic modes. This means that, we recover the presence of waves as seen for the spatio-temporal spectrum for magnetic field fluctuations. In the case of the slow branch, energy accumulates mostly in modes with low $k_y$ and low $\omega/B_0$, typically $k_z<40$ and $\omega/B_0<10$ and increasing with the value $f_0$. We also measure the energy near the wave modes using Eq. \eqref{cuantificacion}. \citet{Spangler2003} presents some theoretical results on the relation between density and magnetic field fluctuations. For instance, \citet{Bhattarcherjee1998} showed that if the background plasma had zero order spatial gradients, $\delta\rho/\overline{\rho}\sim\delta b/B_0$. This theoretical work was inspired by a number of investigations which have utilized the Helios spacecraft dataset to study the dependence of pressure or density fluctuations on velocity and magnetic field fluctuations \citep{tu1994,Bruno1995a,Bavassano1995b,Klein1993}. In the present paper, we will focus on analysing the amount of energy located around wave modes for density and magnetic field fluctuations as shown in Figure \ref{cuanti_b0_th_2_bx_055}. A similar amount of energy was found for both density and magnetic field fluctuations. In this case, the energy reaches a maximum for $k_y<20$. Although there might be a relation between these two quantities due to their similar energetic behaviours, we plan to use different methods that allow us to more accurately compare density and magnetic field fluctuations in a future work. 

\section{Conclusion}\label{sec:conclusion}

We investigated the interplay between wave modes and turbulence under the compressible MHD approach, for small and moderate spatial resolution DNSs. We used histograms to study density fluctuations for different values of Mach number, the compressibility of the forcing and mean magnetic field, and found that there are larger fluctuations as we increase these parameters. In particular, when we turn on the magnetic field the variance (or density fluctuation value) is increased by about an order of magnitude. Then, we analysed different types of fluctuations present in the spatial spectrum and found important differences in the shape of the spectra while increasing the Mach number. Nevertheless, the wave number scaling for the spatial spectra was still compatible with theoretical predictions. We also found that the energy transfer is dominated by incompressible or Alfvénic fluctuations. 

We  used spatio-temporal spectra of different magnetic field components to study the presence of waves in compressible MHD turbulence, varying the Mach number and the mean magnetic field. We observed that, with the magnetic field fixed, the energy located around a given compressible wave mode increased as the Mach number approached to one (and therefore speeds are closer to the speed of sound). When we vary the magnetic field, with Mach number fixed, we go from having most of the energy spread across the spectrum to being located around fast and slow modes, due to the turbulent dynamics of the system and the growth of anisotropic turbulence. In all cases, we detected the presence of linear waves in the particular spatio-temporal spectrum that we have studied. For parallel fluctuations, Alfvén waves dominated the linear dynamics of the system. Although magnetosonic waves present low energy content, as predicted by the weak wave turbulence theories \citep{chandran2005,chandran2008}, they are not negligible, as we could see from the case of the spatio-temporal spectrum $E_{xx}(k_x=0,k_y=15,k_z)$. Whereas, for perpendicular fluctuations, the energy accumulated with a non-negligible fraction at high frequencies around the fast magnetosonic branch and at low frequencies near the modes with $\omega/B_0=0$ (slow modes or the non-propagating modes). Thus, fast waves may have a role in the dynamics with implications for particle acceleration and other physical processes in the solar wind. 

Finally, we investigated the spatio-temporal spectra of density fluctuations and recovered the presence of waves as seen for the magnetic field fluctuations. We quantify the energy along the dispersion relations of the different waves modes using an integral method and compared with the ones obtained from magnetic field fluctuations. Although we found similar energetic behaviour, different methods that allow us to compare these two quantities will be addressed in future studies. Moreover, in forthcoming works, we plan to compare these numerical results with $\textit{in situ}$ data provided by space missions related to the study of waves and turbulence in the solar wind as the recent Parker Solar Probe \citep{Fo2016} and Solar Orbiter \citep{Mu2020}. Nevertheless, the present study is an important first step for understanding the dynamics of compressible plasma flows and the interplay between waves, turbulence and the impact of the forcing in the plasma.

\section{Acknowledgment}

M.B., N.A. and P.D. acknowledge financial support from CNRS/CONICET Laboratoire International Associé (LIA) MAGNETO. N.A. acknowledges financial support from the following grants: PICT 2018 1095 and UBACyT 20020190200035BA. P.D. acknowledges financial support from the following grants: PIP CONICET 11220150100324CO,and  PICT 2018-4298.

%\bibliography{cites}{}
\bibliographystyle{aasjournal}

\end{document}